\documentclass[preprint,12pt]{elsarticle}

\usepackage{amssymb}

\journal{Journal of Network and Computer Applications}

\begin{document}

\begin{frontmatter}



\title{A lightweight dynamic pseudonym identity based authentication and key agreement protocol without verification tables for multi-server architecture}


\author{Kaiping Xue, Peilin Hong, Changsha Ma}

\address{The Information Network Lab of EEIS Department, USTC, Hefei, 230027, China}

\begin{abstract}
Traditional password based authentication schemes are mostly considered in single server environments. They are unfitted for the multi-server environments from two aspects. On the one hand, users need to register in each server and to store large sets of data, including identities and passwords.  On the other hand, servers are required to store a verification table containing user identities and passwords. Recently, On the base on Sood et al.'s protocol(2011), Li et al. proposed an improved dynamic identity based authentication and key agreement protocol for multi-server architecture(2012). Li et al. claims that the proposed scheme can make up the security weaknesses of Sood et al.'s protocol. Unfortunately, our further research shows that Li et al.'s protocol contains several drawbacks and can not resist some types of known attacks, such as replay attack, Deny-of-Service attack, internal attack, eavesdropping attack, masquerade attack, and so on. In this paper, we further propose a light dynamic pseudonym identity based authentication and key agreement protocol for multi-server architecture. In our scheme, service providing servers don't need to maintain verification tables for users. The proposed protocol provides not only the declared security features in Li et al.'s paper, but also some other security features, such as traceability and identity protection.

\end{abstract}


\end{frontmatter}


\section{Introduction}
\label{}
With the rapid growth of modern computer networks, increasing numbers of systems contain a certain quantity of service providing servers around the world and provide services via the Internet. It's important to verify the legitimacy of a remote user in a public environment before he/she can access the service. But traditional password based authentication schemes are mostly considered in single server environments. They are unfitted for the multi-server environments from two aspects. On the one hand, users need to register in each server and to store large sets of data, including identities and passwords.  On the other hand, servers are required to store a verification table containing user identities and passwords. \cite{Hwang-Li} firstly proposed a remote authentication scheme using smart card based on Elgamal's public key cryptosystem\cite{Elgamal}, which doesn't need to maintain verification tables. After that, numerous smart card based single-server authentication schemes using one-way hash functions had been proposed\cite{passwd-based-single1,passwd-based-single2, passwd-based-single3, passwd-based-single4,passwd-based-single5, passwd-based-single6,passwd-based-single7}. However, it is still hard for a user to use different smart cards to login and access different remote servers. This is because users still need to remember numerous sets of identities and passwords. In order to resolve this problem, several schemes have been proposed to the study of authentication and key agreement in the multi-server environment\cite{passwd-based-multi1,passwd-based-multi2,passwd-based-multi3,passwd-based-multi4,passwd-based-multi5,passwd-based-multi6, passwd-based-multi7}, all of which claim not to store verification tables. Most of these schemes can be divided into three categories: hash-based, symmetric cryptosystem based and public-key cryptosystem based. Hash-based protocols are considered to be with the most efficiency.

Among these schemes, in 2009, Hsiang and Shih proposed a dynamic identity and one-way hash based remote user authentication protocol for multi-server architecture without a verification table\cite{passwd-based-multi1}. However, in 2011, Sood et al.\cite{passwd-based-multi2} pointed that Hsiang and Shih's protocol can not resist many types of security attacks, such as replay attack, impersonation attack and stolen smart card attack. Then Sood et al. proposed an improved scheme which is claimed to achieve user anonymity and resist different types of common security attacks. Recently, in \cite{passwd-based-multi7}, Li et al. found that Sood et al.'s protocol is still vulnerable to some types of known attacks, such as replay attack, stolen smart card attack and so on. Also the mutual authentication and key agreement phase of Sood et al.'s protocol can not be successfully finished within some specific scenes. Furthermore, in \cite{passwd-based-multi7}, they proposed an improved dynamic identity based authentication and key agreement protocol for multi-server architecture, which is claimed to remove the aforementioned weaknesses of Sood et al.'s protocol. Unfortunately, our further research shows that Li et al.'s protocol contains several drawbacks and can not resist some types of known attacks, such as leak-of-verifier attack, stolen smart card attack, eavesdropping attack, replay attack, deny-of-service attack and forgery attack and so on.

The rest of this paper is organized as follows: Section 2 gives the overview of Li et al.'s protocol; Section 3 points out the security weaknesses of the protocol in details. Section 4 gives our proposed protocol. Security and performance analysis of our proposed protocol are given in Section 5 and Section 6. At last, Section 7 presents the overall conclusion. 	

\begin{table}[!ht]
\caption{Notations used in Li et. al.'s paper}
\centering
 \resizebox{1.0\columnwidth}{!}{
\begin{tabular}{l|l}
\hline
$U_i$& a user\\
$S_j$& a service providing server\\
$CS$& the control server\\
$ID_i$& the identity of $U_i$\\
$SID_j$& the identity of $S_j$\\
$x$& the master secret key\\
$y$& the secret number\\
$b$& a random number chosen by the user for registration\\
$CID_i$& the dynamic identity generated by $U_i$ for authentication\\
$SK$ & session key shared among the user, the server and $CS$\\
$N_{i1}$, $N_{i2}$, $N_{i3}$ & random numbers chosen by $U_i$, $S_j$ and $CS$\\
$h(\cdot)$& a one way hash function \\
$\oplus$& the bitwise XOR operation\\
$||$ & the bitwise concatenation operation \\
\hline
\end{tabular}}
\end{table}

\section{Overview of Li et al.'s protocol}
In this section, we give the overview of Li et al.'s proposed protocol, which is an enhanced scheme from Sood et al.'s protocol. We firstly summarize the notations used through out Li et al.'s paper in Table 1. Li et al.'s protocol involves 3 kinds of participants: users(taking $U_i$ for example), service providing servers(taking $S_j$ for example), and the control server($CS$). $CS$ is a trusted third party responsible for the registration and authentication of the users and the service providing servers. $CS$ chooses two security elements $x$ and $y$.In the registration phase, $S_j$ obtains $h(SID_j||y)$ and $h(x||y)$ from $CS$ via a secure channel. $U_i$ randomly selects a number $b$, and computes $A_i=h(b||P_i)$. After the initialization and the registration phases, $U_i$ can get a smart card from $CS$ via a secure channel. The following elements, $h(\cdot)$, $h(y)$ and $b$ are stored in the smart card for the user $U_i$:
\begin{eqnarray}
\left.\begin{array}{l}
C_i=h(ID_i||h(y)||A_i)\\
D_i=B_i\oplus h(ID_i||A_i)=h(ID_i||x)\oplus h(ID_i||A_i)\\
E_i=B_i\oplus h(y||x)=h(ID_i||x)\oplus h(y||x)
\end{array} \right.
\end{eqnarray}

\begin{figure}[!htb]
\begin{center}
\includegraphics[width=0.60\textwidth]{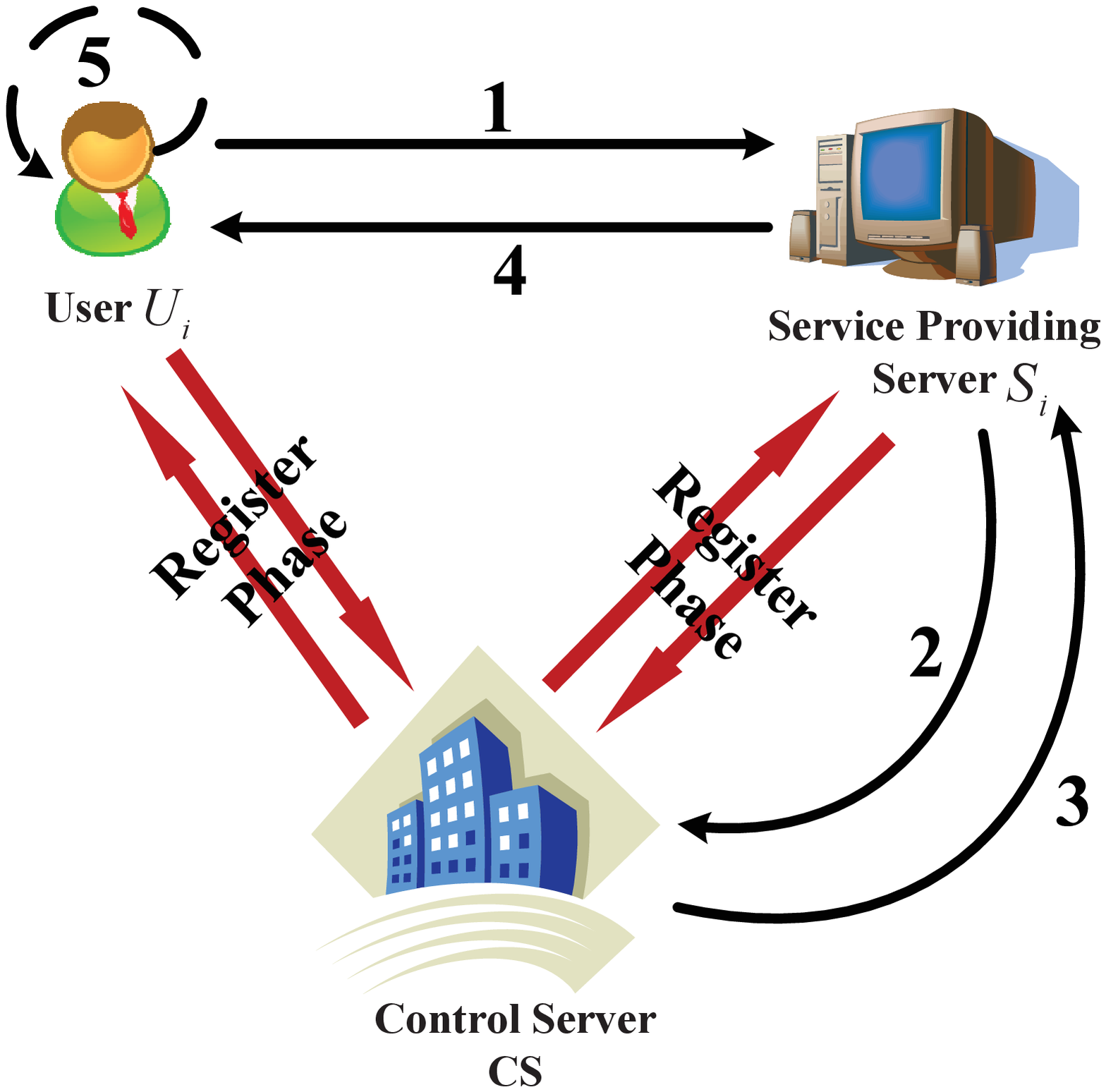}
\caption{Demonstration of Register, Authentication and key agreement phases of Li et al.'s protocol }
\end{center}
\end{figure}

In $U_i$'s login phase, $U_i$ inserts his smart card into a terminal and inputs his identity $ID_i$ and password $P_i$, then computes $A^*_i=h(b||P_i)$ and $C^*_i=h(ID_i||h(y)||A^*_i)$. If $C^*_i$ is equal to the stored $C_i$, $U_i$ is considered as a legitimate user. Else, the terminal rejects $U_i$'s login request.  After the verification, the authentication and key agreement phase takes place among $U_i$, $S_j$ and $CS$, as depicted in Figure 1. We introduce them as follows:

\begin{enumerate}[Step 1:]
\item\footnote{In the description of \cite{passwd-based-multi7}, except for sending the message, this step is included in the login step.} \textbf{$U_i$ $\rightarrow$ $S_j$: \{$F_i$, $G_i$, $P_{ij}$, $CID_i$\}}.\\
$U_i$ computes $B_i=D_i\oplus h(ID_i||A_i)$ and generates a random number $N_{i1}$. Then $U_i$ computes  $F_i$, $G_i$, $P_{ij}$, $CID_i$ as follows:
\begin{eqnarray}
\left.\begin{array}{l}
F_i=h(y)\oplus N_{i1}\\
G_i=h(B_i||A_i||N_{i1})\\
P_{ij}=E_i\oplus h(h(y)||N_{i1}||SID_j)\\
CID_i=A_i\oplus h(B_i||F_i||N_{i1})
\end{array} \right.
\end{eqnarray}
Then, $U_i$ sends \{$F_i$, $G_i$, $P_{ij}$, $CID_i$\}to $S_j$ over a public channel.
\item \textbf{$S_j$ $\rightarrow$ $CS$: \{$F_i$, $G_i$, $P_{ij}$, $CID_i$, $SID_j$, $K_i$, $M_i$ \}}.\\
After receiving the message from $U_i$, the server $S_j$ randomly selects a number $N_{i2}$ and computes $K_i$, $M_i$ as follows:
\begin{eqnarray}
\left.\begin{array}{l}
K_i=h(SID_j||y)\oplus N_{i2}\\
M_i=h(h(x||y)||N_{i2})
\end{array} \right.
\end{eqnarray}
Then $S_j$ sends \{$F_i$, $G_i$, $P_{ij}$, $CID_i$, $SID_j$, $K_i$, $M_i$ \} to $CS$ over the public channel.
\item \textbf{$CS$ $\rightarrow$ $S_j$: \{$Q_i$, $R_i$, $V_i$, $T_i$ \}}.\\
After receiving the message from $S_j$, $CS$ gets $N_{i2}=K_i\oplus h(SID_j||y)$ and $M^*=h(h(x||y)||N_{i2})$. Then $CS$ verifies whether $M^*$ is equal to the received $M_i$. If not, $CS$ terminates the session; Else, the legitimacy of $S_j$ is verified by $CS$. After that, $CS$ computes the following elements:
\begin{eqnarray}
\left.\begin{array}{l}
N_{i1}=F_i\oplus h(y)\\
B_i=P_{ij}\oplus h(h(y)||N_{i1}||SID_j)\oplus h(y||x)\\
A_i=CID_i\oplus h(B_i||F_i||N_{i1})\\
G^*_i=h(B_i||A_i||N_i1)
\end{array} \right.
\end{eqnarray}
Then $CS$ verifies whether $G^*$ is equal to the received $G_i$. If not, $CS$ terminates the session; Else, the legitimacy of $U_i$ is verified by $CS$. $CS$ randomly selects a number $N_{i3}$, and computes the following elements:
\begin{eqnarray}
\left.\begin{array}{l}
Q_i=N_{i1}\oplus N_{i3}\oplus h(SID_j||N_{i2})\\
R_i=h(A_i||B_i)\oplus h(N_{i1}\oplus N_{i2}\oplus N_{i3})\\
V_i=h(h(A_i||B_i)||h(N_{i1}\oplus N_{i2}\oplus N_{i3}))\\
T_i=N_{i2}\oplus N_{i3}\oplus h(A_i||B_i||N_{n1})
\end{array} \right.
\end{eqnarray}
Then $CS$ sends \{$Q_i$, $R_i$, $V_i$, $T_i$ \}to $S_i$ over a public channel.
\item \textbf{$S_j$ $\rightarrow$ $U_i$: \{$V_i$, $T_i$\}}.\\
After receiving the message from $CS$, $S_j$ computes:
\begin{eqnarray}
\left.\begin{array}{l}
N_{i1}\oplus N_{i3}=Q_i \oplus h(SID_j||N_{i2})\\
h(A_i||B_i)=R_i\oplus h(N_{i1}\oplus N_{i3}\oplus N_{i2})\\
V^*_i=h(h(A_i||B_i)||h(N_{i1}\oplus N_{i3}\oplus N_{i2}))
 \end{array} \right.
\end{eqnarray}
Then $S_j$ verifies whether $V^*_i$ is equal to the received $V_i$. If not, $S_j$ terminates the session; Else, the legitimacy of $CS$ is verified by $S_j$. After that, $S_j$ sends the message \{$V_i$, $T_i$\} to $U_i$.
\item
After receiving the message from $S_j$, $U_i$ computes to get $V'_i$ as follows:
\begin{eqnarray}
\left.\begin{array}{l}
N_{i2}\oplus N_{i3}=T_i \oplus h(A_i||B_i||N_{i1})\\
V'_i=h(h(A_i||B_i)||h(N_{i2}\oplus h(N_{i3})\oplus h(N_{i1})))
 \end{array} \right.
\end{eqnarray}
Then $U_j$ verifies whether $V'_i$ is equal to the received $V_i$. If not, $U_i$ terminates the session; Else, the legitimacy of $CS$ and $S_j$ is verified by $U_i$.
\end{enumerate}

 Finally, $U_i$, $S_j$ and $CS$ can separately compute the shared session key $SK$ as follow:
\begin{eqnarray}
SK=h(h(A_i||B_i)||(N_{i1}\oplus N_{i2}\oplus N_{i3}))
\end{eqnarray}

\section{Security weakness analysis of the protocol}
Although in \cite{passwd-based-multi7}, the authors claimed that their protocol can resist many types of security attacks. Unfortunately, our further research shows that Li et al.'s protocol contains several drawbacks and can not resist some types of known attacks, such as replay attack, deny-of-service attack, smart card forgery attack, eavesdrop attack and forgery attack. The analysis in details is described as follows.

\subsection{Replay attack and Deny-of-Service attack}
Assume that a malicious attacker can eavesdrop the first sending message from a legitimate user to the server $S_k$ in Step1 of the authentication and key agreement phase. If the message \{$F_i$, $G_i$, $P_{ij}$, $CID_i$\} is eavesdropped,  replay attacks can easily be launched by retransmitting \{$F_i$, $G_i$, $P_{ij}$, $CID_i$\} to $S_j$. This type of attacks can trick the server $S_k$ and $CS$ into implementing the following steps Step2-4. Moreover, $S_K$ and $CS$ can not identify the message replayed by the malicious attackers. Even if the user cannot get the final correct session key $SK$, the server $S_k$ and $CS$ have made great consumption of computing resources, communication resources and storage resources. A large number of replay attacks launched at the same time will form a Deny-of-Service attack, which prevents normal visits from legitimating legitimate users.

\subsection{Internal attack}
Assume there is an inside malicious user who has a legitimate smart card. From the elements stored in the smart card, the malicious user can straightly get $h(y)$. The malicious attacker $U_f$ can firstly compute his/her $B_f$($ = D_f \oplus h(ID_f||A_f)$), and then computes $h(y||x)=E_f\oplus B_f$. By Knowing $h(y)$ and $h(y||x)$, the attacker can further launch eavesdrop attacks to get the session key shared among any other users, the related service providing servers and $CS$.

\subsection{Smart card forgery attack}
Li et al.'s protocol lacks of verification of $A_i$ and $B_i$ by $CS$, thus a malicious attacker known $h(y)$ and $h(y||x)$ in advance can arbitrarily forge a new smart card. If the attacker wants to forge $U_s$'s smart card, he/she firstly sets $A_s=Num1$ and $B_i=Num2$, where $Num1$ and $Num2$ are two random numbers with the same length as $A_i$, $B_i$. The elements of a forgery smart card can be further set as:
\begin{eqnarray}
\left.\begin{array}{l}
C_s=h(ID_s||h(y)||A_s)=C_s=h(ID_s||h(y)||Num1)\\
D_s=B_s\oplus h(ID_s||A_s)=Num2\oplus h(ID_s||Num1)\\
E_s=B_s\oplus h(y||x)=Num2\oplus h(y||x)
 \end{array} \right.
\end{eqnarray}

Then if the malicious attacker wants to access the service providing server $S_j$ by using this forgery smart card. The first message can be computed as:
\begin{eqnarray}
\left.\begin{array}{l}
F_s=h(y)\oplus N_{s1}\\
G_s=h(B_s||A_s||N_{s1})=h(Num2||Num1||N_{s1})\\
P_{sj}=E_s\oplus h(h(y)||N_{s1}||SID_j)=Num2\oplus h(y||x)\oplus h(h(y)||N_{s1}||SID_j)\\
CID_s=A_s\oplus h(B_s||F_s||N_{s1})=Num1\oplus h(Num2||F_s||N_{s1})
 \end{array} \right.
\end{eqnarray}

Following Li et al.'s protocol, this message can successfully pass the legitimacy verification by $CS$ and $S_j$. If the random numbers separately chosen by $S_j$ and $CS$ are $N_{s2}$ and $N_{s3}$, the malicious attacker, $S_j$ and $CS$ can successfully agree on a common session key $SK=h(h(Num1||Num2)||(N_{s1}\oplus N_{s2}\oplus N_{s3}))$.

\subsection{Eavesdropping attack}
Assume the authentication and key agreement phase takes place among the legitimate user $U_m$, the service providing server $S_n$ and the control server $CS$.

There is a malicious attacker who has the ability of eavesdropping all of the messages exchanged among these three participants. Furthermore, The malicious attacker is assumed to have known $h(y)$, $h(y||x)$ in advance. The first message is  \{$F_m$, $G_m$, $P_{mn}$, $CID_m$\} send from $U_m$. From $F_m$, $N_{m1}$ can been easily obtained as follow:
\begin{eqnarray}
N_{m1}=h(y)||F_m
\end{eqnarray}
Next, $E_m$ can be extracted from $P_{mn}$, then $B_m$ can be extracted from $E_m$. The details are described as follows:
\begin{eqnarray}
\left.\begin{array}{l}
E_m=P_{mn}\oplus h(h(y)||N_{m1}||SID_n)\\
B_m=E_m\oplus h(y||x)
 \end{array} \right.
\end{eqnarray}
After that from $CID_m$, $A_m$ can also be easily extracted as:
\begin{eqnarray}
A_m=CID_m\oplus h(B_m||F_m||N_{m1})
\end{eqnarray}

From the above process, only a sending message via a public channel can leak crucial security information ($A_m$, $B_m$, $N_{m1}$) of $U_m$. Also $E_m$ stored in $U_m$'s smart card can also be got. Although because of the user anonymity support, the malicious attacker can not obtain $U_m$'s identity $ID_m$ to compute $C_m$ and $D_m$, but next we will describe how to extract the final session key $SK$.

After eavesdropping the message send in Step3 or Step4. the malicious attacker can extract $N_{m2}\oplus N_{m3}$ from $T_m$ as follow:
\begin{eqnarray}
N_{m2}\oplus N_{m3}=T_m\oplus h(A_m||B_m||N_{i1})
\end{eqnarray}

Now, the malicious attacker can compute the final session key negotiated among $U_m$, $S_n$ and $CS$. Furthermore, he/she can decrypted all the encrypted data between $U_m$ and $S_n$.

\subsection{Masquerade attack to pose as a legitimate user}
After successfully obtaining security information of a legitimate user(such as $U_m$) via the eavesdrop attack described in Section 3.4, The attacker can launch the masquerade attack to act as the legitimate user. By means of the internal attack, the malicious attackers can know $h(y)$ and $h(y||x)$. By means of the eavesdrop attack, the malicious attacker can further compute $A_m$, $B_m$ and $E_m$. By virtue of these information, the malicious attacker can pose as $U_m$ to launch authentication and key agreement phase to any other service providing server(Take $S_p$ for example) and $CS$.

Firstly, the malicious attacker randomly select a number $N_{MA}$ and can successfully forge the first step message to pretend to be $U_m$:
\begin{eqnarray}
\left.\begin{array}{l}
F_m=h(y)\oplus N_{MA}\\
G_m=h(B_m||A_m||N_{MA})\\
P_{mp}=E_m\oplus h(h(y)||N_{MA}||SID_p)\\
CID_m=A_m\oplus h(B_m||F_m||N_{MA})
 \end{array} \right.
\end{eqnarray}

Then assume $S_p$ and $CS$ separately select random numbers $N_{m2}$ and $N_{m3}$, and Step2-Step4 are performed normally. Then the malicious attacker, $S_j$ and $CS$ ``successfully'' agree on a session key $SK=h(h(A_m||B_m)||(N_{MA}\oplus N_{m2}\oplus N_{m3}))$. But unfortunately $S_p$ and $CS$ mistakenly believe that they are communicating with the legitimate user $U_m$.

\subsection{Masquerade attack to pose as a legitimate service providing server}
First assume that the malicious attacker has eavesdropped a message send from $S_n$ to get $K_i$ and $M_i$. Furthermore assume a legitimate user $U_m$'s security information has been leaked to the malicious attacker based on the internal attack and the eavesdrop attack. When $U_m$ wants to login the server $S_n$, he/she selects a random number $N_{m1}$ and sends the first message in Step1(\{$F_m$, $G_m$, $P_{mn}$, $CID_m$\}) to the service providing server $S_n$. The malicious attacker can attack the real server $S_n$ to be down and masquerades to be $S_n$ himself/herself. After eavesdropping this message, the malicious attacker can attach $K_i$ and $M_i$ in the first message:\{$F_m$, $G_m$, $P_{mn}$, $CID_m$, $SID_n$, $K_i$, $M_i$ \}. This message can also successfully pass $CS$'s verification. $N_{m3}$ is the random number selected by $CS$. After implementing of Step3 and Step4, the user $U_m$ and $CS$ can compute the session key as
\begin{eqnarray}
SK=h(h(A_m||B_m)||h(N_{m1}\oplus N_{i2}\oplus N_{m3}))
\end{eqnarray}
And unfortunately $U_m$ mistakenly believe that he/she is communicating with the legitimate true $S_n$. Although the malicious attacker can not extract the random number $N_{i2}$ from $K_i$, he/she still can exact the session key $SK$ by means of ``masquerade attack as a legitimate user'' described in Section 3.5. So the malicious attacker can not only masquerade to be the real server, but also decrypt the encrypted data send from the user in the dark.

\begin{figure}[!htb]
\begin{center}
\includegraphics[width=1.00\textwidth]{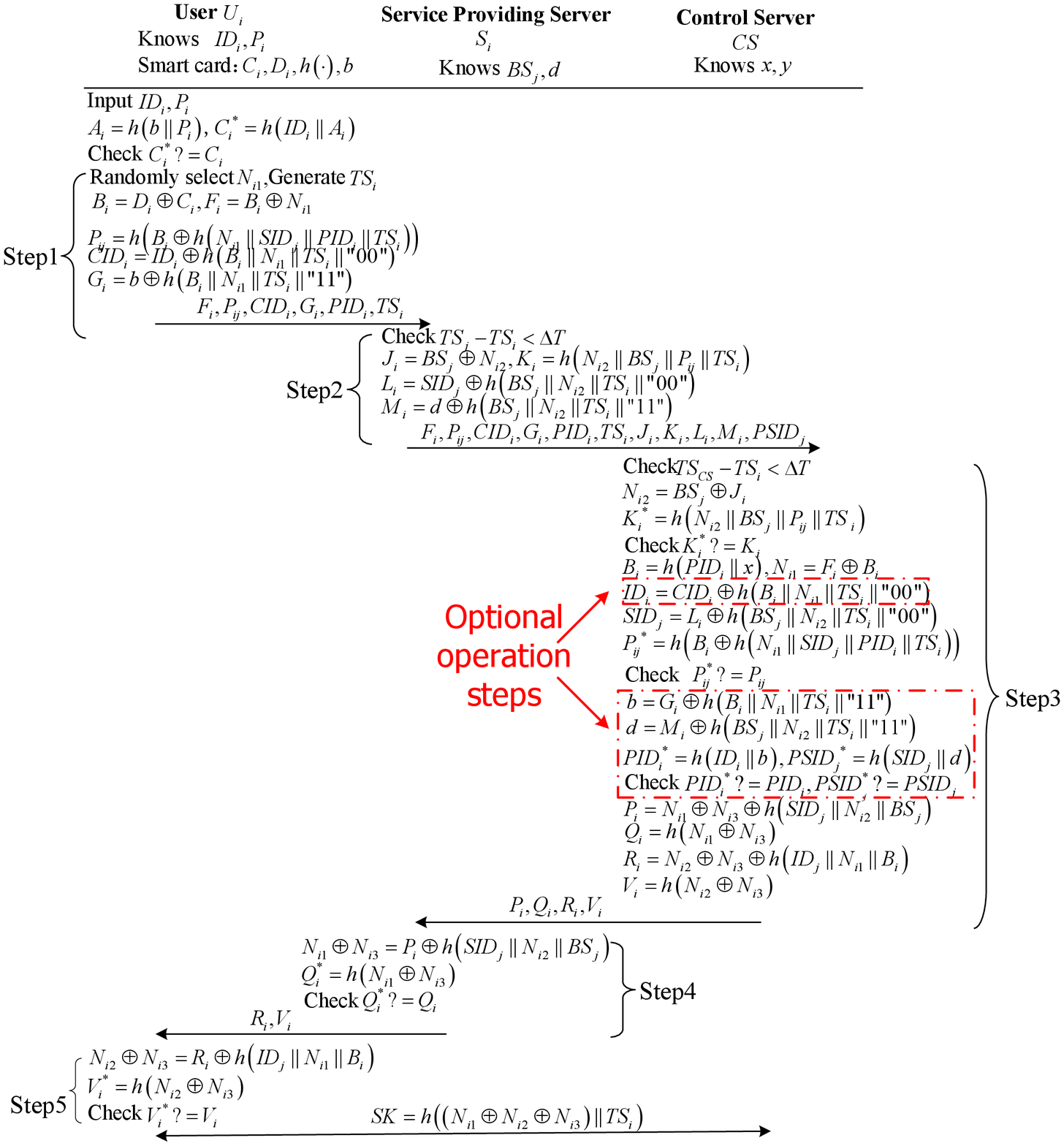}
\caption{The implement phases of our proposed protocol}
\end{center}
\end{figure}

\section{Our proposed improved protocol}
In this section, we will describe an improved protocol to make up the security weaknesses of Li et al.'s protocol. Our protocol contains three kinds of participants(the user, the service providing server and the controlling server) and contains three phases: 1)Initialization and registration phase; 2) login phase; 3)authentication and key agreement phase. Because the notions are different in using from those of Li et al.'s protocol in protocol designing and some new notions are defined, here we firstly give the notations used in our proposed protocol(Summarize in Table 2). We show the protocol in Figure 2 and provide more details as follows.

\begin{table}[!ht]
\centering
\caption{Notations used in our proposed protocol}
 \resizebox{1.0\columnwidth}{!}{
\begin{tabular}{l|l}
\hline
$U_i$& a user\\
$S_j$& a service providing server\\
$CS$& the control server\\
$ID_i$& the identity of $U_i$\\
$SID_j$& the identity of $S_j$\\
$TS_i$& Timestamp value generated by $U_i$\\
$x$& the secret number only known to $CS$\\
$y$& the secret number only known to $CS$\\
$b$& a random number chosen by the user\\
$d$& a random number chosen by the service providing server\\
$PID_i$& the protected pseudonym identity of $U_i$ \\
$PSID_j$& the protected pseudonym identity of $S_j$ \\
$SK$ & session key shared among the user, the server and $CS$\\
$N_{i1}$, $N_{i2}$, $N_{i3}$ & random numbers chosen by $U_i$, $S_j$ and $CS$\\
$h(\cdot)$& a one way hash function \\
$\oplus$& the bitwise XOR operation\\
$||$ & the bitwise concatenation operation \\
\hline
\end{tabular}}
\end{table}

\subsection{Initialization and registration phase}
Assume the control server $CS$ is a trusted third party responsible for registration and authentication of users and service providing servers. $CS$ chooses two random numbers $x$ and $y$.

The registration phase of the user $U_i$ is as follows:

\begin{enumerate}[Step 1:]
\item
The user $U_i$ freely choose his/her identity $ID_i$ and password $P_i$, and randomly choose a number $b$. Then $U_i$ compute $A_i=h(b||P_i)$, and submits the message \{$ID_i$, $b$, $A_i$\} to $CS$ via a secure channel.
\item
After receiving the message, $CS$ first verifies user's legitimacy. Then, $CS$ computes $PID_i=h(ID_i||b)$, $B_i=h(PID_i||x)$. $CS$ sends $B_i$ to $U_i$ via a secure channel.
\item
After receiving the smart card, $U_i$ computes $C_i=h(ID_i||A_i)$ and $D_i=B_i\oplus h(PID_i\oplus A_i)$. Then $U_i$ enters$C_i$, $D_i$, $h(\cdot)$ and $b$ into the smart card. At last, the smart card contains ($C_i$, $D_i$, $h(\cdot)$, $b$).
\end{enumerate}

For the service providing server $S_j$, he/she first chooses a random number $d$, and use his/her identity $S_j$ to register with $CS$. $CS$ computes $PSID_j=h(SID_j||d)$, $BS_j=h(PSD_j||y)$. Then $CS$ sends $BS_j$ to $S_j$ via a secure channel. $S_j$ stores $BS_j$ and $d$ in his/her memory.

\subsection{Login phase}
When the user $U_i$ wants to login to access the server $S_j$, $U_i$ inserts his smart card into a terminal and inputs his/her identity $ID_i$ and password $P_i$, then computes $A^*_i=h(b||P_i)$ and $C^*_i=h(ID_i||A^*_i)$. If $C^*_i$ is equal to the stored $C_i$, $U_i$ is considered as a legitimate user. Otherwise, the terminal rejects $U_i$'s login request.

\subsection{Authentication and key agreement phase}
\begin{enumerate}[Step 1:]
\item \textbf{$U_i$ $\rightarrow$ $S_j$: \{$F_i$, $P_{ij}$, $CID_i$, $G_i$, $PID_i$, $TS_i$\}}.\\
$U_i$ chooses a random number $N_{i1}$ and generates a current Timestamp value $TS_i$. Then $U_i$ computes $B_i$, $F_i$, $CID_i$, $P_{ij}$, $G_i$ as follows:
\begin{eqnarray}
\left.\begin{array}{l}
B_i=D_i\oplus C_i\\
F_i=B_i\oplus N_{i1}\\
P_{ij}=h(B_i\oplus h(N_{i1}||SID_j||PID_i||TS_i))\\
CID_i=ID_i\oplus h(B_i||N_{i1}||TS_i||``00'')\\
G_i=b\oplus h(B_i||N_{i1}||TS_i||``11'')
\end{array} \right.
\end{eqnarray}
Where, ``00'' is a 2-bit binary-``0'', and ``11'' is a 2-bit binary-``1''.

Then, $U_i$ sends \{$F_i$, $P_{ij}$, $CID_i$, $G_i$, $PID_i$, $TS_i$\}to $S_j$ over a public channel.

\item \textbf{$S_j$ $\rightarrow$ $CS$: \{$F_i$, $P_{ij}$, $CID_i$, $G_i$, $PID_i$, $TS_i$, $J_i$, $K_i$, $L_i$, $M_i$, $PSID_j$\}}.\\
After receiving the message from $U_i$, the server $S_j$ first checks whether the session delay is within the tolerable time interval $\Delta T$. Assume the current time is $TS_j$. If $TS_j-TS_i> \Delta T$, the session is timeout and $S_j$ terminates the session; Otherwise, $S_j$ continues to perform the following operations.

$S_j$ randomly selects a number $N_{i2}$ and computes $J_i$, $K_i$, $L_i$, $M_i$ as follows:
\begin{eqnarray}
\left.\begin{array}{l}
J_i=BS_j\oplus N_{i2}\\
K_i=h(N_{i2}||BS_j||P_{ij}||TS_i)\\
L_i=SID_j\oplus h(BS_j||N_{i2}||TS_i||``00'')\\
M_i=d\oplus h(BS_j||N_{i2}||TS_i||``11'')
\end{array} \right.
\end{eqnarray}
Where, `00'' is a 2-bit binary-``0'', and ``11'' is a 2-bit binary-``1''.

Then $S_j$ sends \{$F_i$, $P_{ij}$, $CID_i$, $G_i$, $PID_i$, $TS_i$, $J_i$, $K_i$, $L_i$, $M_i$, $PSID_j$\} to $CS$ over the public channel.

\item \textbf{$CS$ $\rightarrow$ $S_j$: \{$P_i$, $Q_i$, $R_i$, $V_i$ \}}.\\
After receiving the message from $S_j$, $CS$ first checks whether the session delay is within the allow time interval $\Delta T$. Assume the current time is $TS_{CS}$. If $TS_{CS}-TS_i> \Delta T$, the session is timeout and $CS$ terminates the session; $CS$ continues to perform the following operations.

$CS$ computes $BS_j=h(PSID_j||y)$, $N_{i2}=J_i\oplus BS_j$ and $K^*=h(N_{i2}||BS_j||P_{ij}||TS_i)$.  Then $CS$ verifies whether $K^*_i$ is equal to the received $K_i$. If not, $CS$ terminates the session; Otherwise, $CS$ continues to perform the following operations. $CS$ computes the following elements:
\begin{eqnarray}
\left.\begin{array}{l}
B_i=h(PID_i||x)\\
N_{i1}=F_i\oplus B_i\\
ID_i=CID_i\oplus h(B_i||N_{i1}||TS_i||``00'')\\
SID_i=L_i\oplus h(BS_j||N_{i2}||TS_i||``00'')\\
P^*_{ij}=h(B_i\oplus h(N_{i1}||SID_j||PID_i||TS_i))\\
\end{array} \right.
\end{eqnarray}
Then $CS$ verifies whether $P^*_{ij}$ is equal to the received $P_{ij}$. If not, $CS$ terminates the session; Otherwise, $CS$ continues to compute the following elements:
\begin{eqnarray}
\left.\begin{array}{l}
b=G_i\oplus h(B_i||N_{i1}||TS_i||``11'')\\
d=M_i\oplus h(BS_j||N_{i2}||TS_i||``11'')\\
PID^*_i=h(ID_i||b)\\
PSID^*_j=h(SID_j||d)
\end{array} \right.
\end{eqnarray}
Then $CS$ verifies whether $PID^*_i=PID_i$ and $PSID^*_j=PSID_j$. If not, $CS$ terminates the session; Otherwise, $CS$  makes sure the messages are from real $U_i$ and $S_j$. After the verification, $CS$ randomly selects a number $N_{i3}$, and computes $P_i$, $Q_i$, $R_i$ $V_i$ as follows:
\begin{eqnarray}
\left.\begin{array}{l}
P_i=N_{i1}\oplus N_{i3}\oplus h(SID_j||N_{i2}||BS_j)\\
Q_i=h(N_{i1}\oplus N_{i3})\\
R_i=N_{i2}\oplus N_{i3}\oplus h(ID_i||N_{i1}||B_i)\\
V_i=h(N_{i2} \oplus N_{i3})\\
\end{array} \right.
\end{eqnarray}
Then $CS$ sends \{$P_i$, $Q_i$, $R_i$, $V_i$ \}to $S_i$ over a public channel.

\item \textbf{$S_j$ $\rightarrow$ $U_i$: \{$R_i$, $V_i$\}}.\\
After receiving the message from $CS$, $S_j$ firstly computes to get the following elements:
\begin{eqnarray}
\left.\begin{array}{l}
N_{i1}\oplus N_{i3}=P_i \oplus h(SID_j||N_{i2}||BS_j)\\
Q^*_i=h(N_{i1}\oplus N_{i3})
 \end{array} \right.
\end{eqnarray}
Then $S_j$ verifies whether $Q^*_i$ is equal to the received $Q_i$. If not, $S_j$ terminates the session; Otherwise, the legitimacy of $CS$ is verified by $S_j$. After that, $S_j$ sends the message \{$R_i$, $V_i$\} to $U_i$.

\item
After receiving the message from $S_j$, $U_i$ computes to get $V^*_i$ as follows:
\begin{eqnarray}
\left.\begin{array}{l}
N_{i2}\oplus N_{i3}=R_i \oplus h(ID_i||N_{i1}||B_i)\\
V^*_i=h(N_{i2}\oplus N_{i3})
 \end{array} \right.
\end{eqnarray}
Then $U_j$ verifies whether $V^*_i$ is equal to the received $V_i$. If not, $U_i$ terminates the session; Otherwise, the legitimacy of $CS$ and $S_j$ is verified by $U_i$.

\end{enumerate}

Finally, $U_i$, $S_j$ and $CS$ can separately compute the common session key $SK$ as follow:
\begin{eqnarray}
SK=h((N_{i1}\oplus N_{i2}\oplus N_{i3})|| TS_i))
\end{eqnarray}

\subsection{password updating phase}
After password based verification in the registration phase, the user $U_i$'s password $P_i$  does not appear in $B_i$. Thus password updating/changing can happen in anytime. $U_i$ need to submit his/her $ID_i$ and $A'_i$ with new password $P'_i$ to $CS$ via a secure channel. $CS$ updates $U_i$'s password in its verification table. Meanwhile, $U_i$ can update the parameters in his/her smart card:
\begin{eqnarray}
\left.\begin{array}{l}
C'_i=h(ID_i||A'_i)\\
D'_i=B_i\oplus h(PID_i\oplus A'_i)
 \end{array} \right.
\end{eqnarray}

\subsection{dynamic identity updating phase}
In order to prevent malicious attackers linking eavesdropped messages of different sessions, we can update the user's $PID$ periodically to provide security. $U_i$ reselects a random number $b^\#$, and compute $A^\#_i=h(b^\#||P_i)$. Then $U_i$ submits \{$ID_i$,$b^\#$, $A^\#_i$\} to $CS$. After verifying $U_i$'s legitimacy, $CS$ recomputes $PID^\#_i=h(ID_i||b^\#)$, $B^\#_i=h(PID^\#_i||x)$ and submits $B^\#_i$ to $U_i$ via a secure channel. After receiving $B^\#_i$, $U_i$ computes $C^\#_i=h(ID_i||A^\#_i)$, $D^\#_i=B^\#_i\oplus h(PID^\#_i\oplus A^\#_i)$. At last the smart card is updated to \{$C^\#_i$, $D^\#_i$,$h(\cdot)$, $b^\#$\}. Now $U_i$'s protected pseudonym identity $PID_i$ is dynamically changed to $PID^\#_i$.

Service providing servers can also periodically update their protected pseudonym identities. Take $S_j$ for example, $S_j$ reselects a random number $d^\#$, and use his/her identity $S_j$ to register with $CS$. $CS$ computes $PSID^\#_j=h(SID_j||d^\#)$, $BS^\#_j=h(PSD^\#_j||y)$. Then $CS$ sends $BS^\#_j$ to $S_j$ via a secure channel. $S_j$ updates $BS^\#_j$ and $d^\#$ in his/her memory.

\section{Security analysis of our protocol}
In this section, we summarize security analysis of our proposed protocol and compare it with other two related protocols. First we list security functionality comparison among our protocol and other two related protocols in Table 3. It demonstrates that our protocol is more secure than other two related protocols.

\begin{table}[h]
\centering
\caption{Security functionality comparison of our protocol and two other related protocols}
 \resizebox{1.0\columnwidth}{!}{
\begin{tabular}{llll}
\hline
Security & Our proposed& Li et al.'s&Sood et al.'s\\
functionality&protocol&protocol(2012)&protocol(2011)\\
\hline
User anonymity& Yes &Yes &Yes\\
Mutual authentication & Yes &Yes &Yes\\
Session key agreement & Yes &Yes &Yes\\
Password updating & Yes &Yes &Yes\\
Dynamic identity updating &Yes &No &No\\
Traceability &Yes &No &No\\
Identity protection &Yes &No &No\\
Resistance of Insider attack &Yes &No &No\\
Resistance of Stolen smart card attack &Yes &Yes &No\\
Resistance of replay attack &Yes &No &No\\
Resistance of Deny-of-Service attack &Yes &No &No\\
Resistance of eavesdrop attack &Yes &No &No\\
Resistance of masquerade attack &Yes &No &No\\
\hline
\end{tabular}}
\end{table}

Here we discuss the main security features of our proposed protocol in details:

\subsection{Providing user anonymity}

For the user $U_i$, we use $PID_i$ instead of $ID_i$. By using protected pseudonym identities of users instead of real ones, the malicious attacker can not get user identities. Meanwhile service providing servers can not know users' real identities either. In this way, our protocol provides user anonymity. Furthermore, updating users' pseudonym identities periodically can prevent the malicious attacker linking eavesdropped messages of different sessions from the same user.

\subsection{Providing traceability}
Despite of user anonymity, $CS$ can still extract users' real identities and link them with protected pseudonym identities. This make our protocol have the feature of traceability. This is newly-added function in our proposed protocol different from Li et al.'s protocol.

\subsection{Providing identity protection}
Using protected pseudonym identities of users and service providing servers ensures that only legitimate $CS$ can get their real identities. This can prevent the leakage of private user identities and server identities to malicious attackers. Moreover, in order to prevent malicious attackers link eavesdropped messages of different sessions, protected pseudonym identities of users and service providing servers are dynamic and can changed in any time.

\subsection{Resistance of insider attack and smart card forgery attack}
As in Section 3.2, within Li et al.'s protocol, an internal attack can cause information leakage. $h(y)$ and $h(y||x)$ are the common parameters for all users, which can further launch eavesdrop attacks, smart card forgery attacks, masquerade attacks and so on. In our proposed protocol, we do not straightly use $h(y)$, $h(x)$, $h(y||x)$ directly. Take the user $U_f$ as insider attacker for example, We use $B_f=h(PID_f||x)$ and compute to get $C_f$, $D_f$ in his/her smart card. $U_f$ can not guess to generate parameters of any other users' smart cards and can not masquerade as any other legitimate user by using security information of himself/herself.

\subsection{Resistance of stolen smart card attack}
In our proposed protocol, we firstly assume that if a smart card is stolen, physical protection methods can not prevent malicious attackers to get the stored secure elements. Still take $U_i$ for example, if his/her smart card is stolen, the malicious attacker can get ($C_i$, $D_i$, $h(\cdot)$, $b$). But without inputting right password $P_i$, the malicious attacker can not compute $A_i$, and further extract $B_i$ from $D_i$.

\subsection{Resistance of replay attack and Deny-of-Service attack}
Firstly the timestamp value is used in our proposed protocol which makes the malicious attacker can not use early message to launch replay attacks. This makes replay attacks and Deny-of-Service attacks hard to be launched. Using $P_{ij}$ and $TS_i$ in computing $K_i$ avoids the case in Li et al.'s protocol: If $K_i$ and $M_i$ attached by the service providing server $S_j$ are eavesdropped, they can be used to launch replay attacks, which is described in Section 3.6. Moreover using and verifying timestamp can reduce the success rate of replay attacks.

\subsection{Resistance of eavesdrop attack}
The malicious attacker can not extract private security information from eavesdropping messages over public channels. Different from Li et al.'s protocol, because of using $PID$ in compute $B_i$ and not sharing $h(x)$ and $h(y||x)$ between $CS$ and every user , the malicious attacker can not use one user's elements to extract any other user's security elements in our proposed protocol. Moreover, the malicious attacker can not compute $N_{i1}\oplus N_{i2}\oplus N_{i3}$, so $SK$ can not be computed by the malicious attacker.

\subsection{Resistance of masquerade attack}
The malicious attacker can not derive $U_i$'s security information from eavesdropped sending messages among $U_i$, $S_j$ and $CS$; Meanwhile, the malicious attacker can not forge other user's smart card from known security information of a malicious inside user. Furthermore, Using the timestamp value prevents replay of the first message. Because of the above 3 reasons, users can not be masqueraded by malicious attackers. because of using $P_{ij}$ and $TS_i$ in computing $K_i$, the malicious attacker can not replay $S_j$'s message to attach to  the end of the message in Step 1, thus servers can not be masqueraded by malicious attackers.

\section{Performance Analysis}
In this section, we evaluate the computational complexity, computation overhead, storage overhead of our proposed protocol and give the comparisons with other two related protocols: Li et al.' protocol\cite{passwd-based-multi7} and Sood et al.'s protocol\cite{passwd-based-multi2}. Before analyzing in details, we first give the notation $T_{hash}$ as the time of computing the hash operation. Because $XOR$ and ``$||$'' operations requires very few computations, they are usually omitted in computational complexity computation.
\begin{table}[h]
\centering
\caption{Computational complexity comparison of our protocol and two other related protocols}
 \resizebox{1.0\columnwidth}{!}{
\begin{tabular}{ccccc}
\hline
Protocols &login phase &\multicolumn{3}{c}{authentication and key agreement phase}\\
&$U_i$&$U_i$&$S_j$&$CS$\\
\hline
Our proposed protocol&$2T_{hash}$&$6T_{hash}$&$5T_{hash}$&$8T_{hash}$+$(optional)5T_{hash}$\\
Li et al.'s protocol(2012)&$2T_{hash}$&$8T_{hash}$&$4T_{hash}$&$13T_{hash}$\\
Sood et al.'s protocol(2011)&$1T_{hash}$&$9T_{hash}$&$4T_{hash}$&$11T_{hash}$\\
\hline
\end{tabular}}
\end{table}

Firstly, Computational complexity comparison of our protocol and the other two related protocols is given in Table 4. As in \cite{passwd-based-multi7}, we only take the login phase, authentication and session key agreement phase into consideration. Different from the description in \cite{passwd-based-multi7}, the description of the login phase in Li et al.'s protocol relates only to user legitimacy the by terminal. Similarly, we merge step 2 of the login phase in \cite{passwd-based-multi7} into the first step of the authentication and key agreement phase. The similar decryption modification is adopted to Sood et al.'s protocol\cite{passwd-based-multi2}. Furthermore, There are separately 1 time of hash computation for computing $SK$ for the user, the service providing server and $CS$, which is not mentioned in Table 4. From Table 4, it is obvious that our protocol almost has the same computational complexity with the other two related protocols. In the authentication and key agreement phase of our proposed protocol, $CS$ have five optional hash operations, which proving the function of traceability.

Secondly, we discuss about communication overhead, our proposed protocol and other two related protocols all require 4 times of message transmission in the authentication and key agreement phase. Take $U_i$, $S_j$ and $CS$ for example, four times of message transmission are $U_i\rightarrow S_j$, $S_j\rightarrow CS$, $CS\rightarrow S_j$ and $S_j\rightarrow U_i$, which is demonstrated in Figure 1 .

Thirdly, just as Li et al.'s protocol and Sood et al.'s protocol, our proposed protocol also do not require every service providing server to maintain a verification table. Meanwhile $CS$ maintains a verification table which is only required to search in the registration phase. $CS$ don't need to use the verification table in the authentication and key agreement phase. Each user only needs to have a smart card. Each service providing server(Take $S_j$ for example) only needs to store $BS_j$ and a randomly chosen number$d$ obtained in the registration phase. Besides the verification table, $CS$ only knows $x$ and $y$.

\section{Conclusions}
In this paper, based on discussing the security weaknesses of Li et al.'s protocol, we propose an improved dynamic pseudonym identity based authentication and key agreement protocol, which is suitable for the multi-server environment. Compared with related protocols, our proposed protocol is demonstrated to satisfy all the essential security requirements for authentication and key agreement in the multi-server environment. Meanwhile, in comparison with Li et al.'s protocol and Sood et al's protocol, our proposed protocol keeps efficient, such as low computational complexity, low communication overhead and low storage overhead. In the future, we will survey suitable solutions to further reduce the computational complexity and improve protocol performance while not reducing security.

\section*{Acknowledgements}
This work is supported by the National S\&T Major Project of China under Grant No. 2010ZX03003-002, 2011ZX03005-006, the National Natural Science Foundation of China under Grant No.60903216.

\section*{References}
\bibliographystyle{elsarticle-num}

\end{document}